\documentclass[a4paper,12pt]{article}
\usepackage[dvips]{graphicx}

\begin{document}

\title{Transition Probability to Turbulent Transport Regime} 
\author{
Mitsuhiro Kawasaki, Atsushi Furuya$^{**}$, Masatoshi Yagi, \\Kimitaka Itoh$^{**}$ and Sanae-I. Itoh\\
\\
Research Institute for Applied Mechanics, Kyushu University,\\ Kasuga 816-8580, Japan\\
$^{*}$Interdisciplinary Graduate School of Engineering Sciences,\\ Kyushu University, Kasuga 816-8580, Japan\\
$^{**}$National Institute for Fusion Science, Toki 509-5292, Japan}
\date{}
\maketitle

\begin{abstract}
Transition phenomena between thermal noise state and turbulent
 state observed in a submarginal turbulent plasma are analyzed with
 statistical theory. Time-development of turbulent fluctuation is
 obtained by numerical simulations of Langevin equation which
 contains hysteresis characteristics. Transition rates between two
 states are analyzed. Transition from turbulent state to thermal noise state occurs in entire region between subcritical bifurcation point and linear stability boundary.
\end{abstract}

\section{Introduction}
There have been observed various kinds of formations and destructions of transport barriers. Both in edge and internal regions of high temperature plasmas, the dynamical change often occurs on the short time scale, sometimes triggered by subcritical bifurcation. These features naturally lead to the concept of transition. 

The transition takes place as a statistical process in the presence of statistical noise source induced by strong turbulence fluctuation. As the generic feature the transition occurs with a finite probability when a parameter approaches the critical value. 

The nonequilibrium statistical mechanics, which deals with dynamical phase transitions and critical phenomena, should be extended for inhomogeneous plasma turbulence \cite{Kubo85}. To this end, statistical theory for plasma turbulence has been developed and stochastic equations of motion (the Langevin equations) of turbulent plasma were derived \cite{Itoh99}. The framework to calculate the probability density function (PDF), the transition rates etc. have also been made.

In this paper, we apply the theoretical algorithm to an inhomogeneous
plasma with the pressure gradient and the shear of the magnetic
field. Micro turbulence is known to be subcritically excited from the
thermal noise state \cite{Itoh96}. The transition between thermal noise
state and turbulent state is studied. We show that the transition occurs
stochastically by numerically solving the Langevin equation of the
turbulent plasmas. In order to characterize the stochastic nature of the
transition, the frequency of occurrence of a transition per unit time
(the transition rate) is calculated as a function of the
pressure-gradient and the plasma temperature. The results show that the
transition from the turbulent state to the thermal noise state occurs in
a wide region instead of at a transition point.

\section{Theoretical Framework}
In this section, we briefly review the theoretical framework \cite{Itoh99} used in our analysis of turbulent plasmas. 

The theory is based on the Langevin equation Eq. (\ref{Langevin-k}) derived by renormalizing with the direct-interaction approximation the reduced MHD for the three fields: the electro-static potential, the current and the pressure. 
\begin{equation}
\frac{\partial \mathbf{f}}{\partial t}+\hat{\cal L}\mathbf{f}=\mathbf{\cal N}(t), \ \mbox{where} \ \mathbf{f}(t) \equiv 
\left( \begin{array}{c} \phi(t) \\ J(t) \\ p(t) \end{array} \right).
\label{Langevin-k}
\end{equation}
Since $\mathbf{\cal N}(t)$ is a force which fluctuates randomly in time, the Langevin equation describes the stochastic time-development of the fluctuation of the three fields. 

By analyzing the Langevin equation Eq. (\ref{Langevin-k}), a number of statistical properties of turbulent plasmas can be derived. For example, it was shown that asymptotic forms of the probability distribution functions for the energy of the fluctuation of the electric field obeys a power-law. The analytical formulae of the rate of change of states of plasmas, the transition rates, were also derived. Furthermore, since the renormalized transport coefficients come from the term of the random force $\mathbf{\cal N}(t)$, relations between the fluctuation levels of turbulence and the transport coefficients like the viscosity and the thermal diffusivity were derived.

\section{A Model}
With the theoretical framework briefly described in the previous section, we analyze a model of inhomogeneous plasmas with the pressure-gradient and the shear of magnetic field \cite{Itoh99}. The model is formulated with the reduced MHD of the three fields of the electro-static potential, the current and the pressure. The shear of magnetic field is given as $\mathbf{B}=(0,B_0 s x, B_0)$ where $B_0(x) = \mbox{const}\times (1+\Omega' x+\cdots)$. The pressure is assumed to change in $x-$direction.

It has been known that in this system bifurcation due to the subcritical excitation of the current diffusive interchange mode (CDIM) occurs \cite{Itoh96} as shown in Fig. (1).
\begin{figure}
   \includegraphics[width=12cm,height=8cm,keepaspectratio]{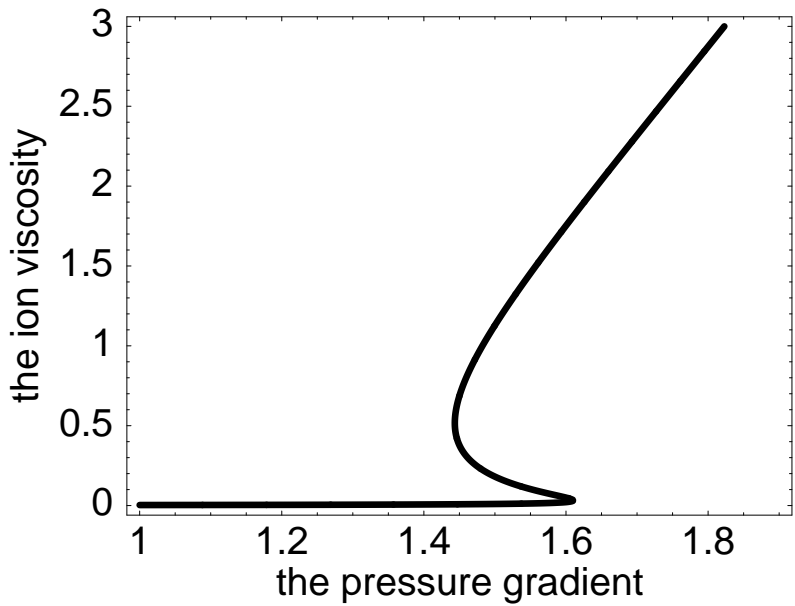}
   \caption{The pressure-gradient dependence of the renormalized ion-viscosity. It is clearly seen that the bifurcation between a low viscosity state (the thermal noise state) and a high viscosity state (the turbulent state) occurs.}
\end{figure}

Figure (1) shows the pressure-gradient dependence of
the turbulent ion-viscosity which is proportional to the fluctuation
level. It is clearly seen that the bifurcation between a low viscosity
state and a high viscosity state occurs. Due to the bifurcation,
transition between the two states and hysteresis are expected to be
observed. We call the low viscosity state ``the thermal noise state'',
since in this state the system fluctuates with thermal noise considered
in the model \cite{Itoh99-2}. We call the high viscosity state ``the
turbulent state'', since the fluctuation level is also large in a strong
turbulent limit \cite{Itoh99}. The ridge point where the turbulent
branch ends is denoted ``the subcritical bifurcation point''. The region
between the subcritical bifurcation point and the ridge near the linear
stability boundary is called ``the bi-stable regime''.

From the deterministic point of view, the transition from the thermal
noise state to the turbulent state is expected to occur at the
ridge point near the linear stability boundary and the transition in the
opposite direction is expected to occur at the subcritical bifurcation point.

\section{Stochastic Occurrence of the Transition}
In order to capture the characteristics of the two states, we concentrate on the time-development of the energy of fluctuation of the electric field, $\varepsilon(t)$. The quantity $\varepsilon(t)$ obeys the coarse-grained Langevin equation Eq. (\ref{Langevin}) which has been derived in \cite{Itoh99}.
\begin{equation}
\frac{d}{dt}\varepsilon(t)=-2 \Lambda(\varepsilon)\varepsilon(t)+\eta(\varepsilon)R(t).
\label{Langevin}
\end{equation}
Here, $R(t)$ is assumed to be the Gaussian white noise. For the detailed formulae of $\Lambda(\varepsilon)$ and $\eta(\varepsilon)$, see \cite{Itoh99-2}. The essential point is that the function $\Lambda(\varepsilon)$ takes both a positive and a negative value in the bi-stable regime. So, the fluctuation of the electric field is suppressed when $\Lambda$ is positive and it is excited when $\Lambda$ is negative. Consequently, there are two metastable states in the bi-stable regime. In addition, $\eta(\varepsilon)$ is a positive function. 

By solving numerically Eq. (\ref{Langevin}), we obtain the following samples of a time serieses. When the pressure-gradient is fixed at the value smaller than the subcritical bifurcation value, as shown in Fig. (2), there is only small fluctuation since the system is always in the thermal noise state. 
\begin{figure}
   \includegraphics[width=10cm,height=5cm]{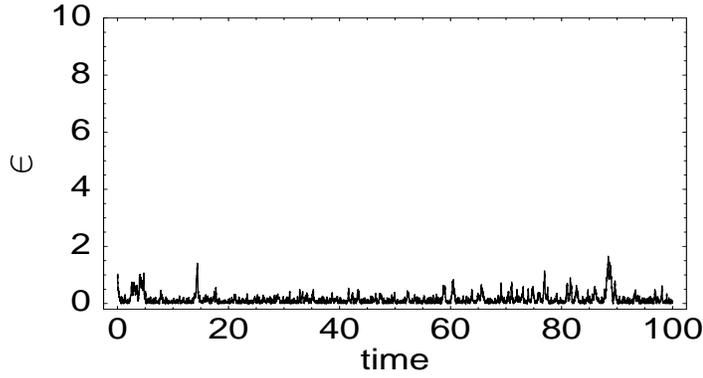}
   \caption{A sample of a time-series of the energy of fluctuation of the electric field $\varepsilon(t)$ when the pressure-gradient is fixed at the value smaller than that of the subcritical bifurcation point. There is only small fluctuation since the system is always in the thermal noise state.}
\end{figure}

On the other hand, when the pressure-gradient takes a value in the bi-stable regime, bursts are observed intermittently as shown in Fig. (3). That is, transition between the thermal noise state and the turbulent state occurs {\it stochastically}. The bursts corresponds to the turbulent state and the laminar corresponds to the thermal noise state. The fact that the residence times at the each states are random leads to the statistical description of the transition with the transition rates described in the next section. 
\begin{figure}
   \includegraphics[width=10cm,height=5cm]{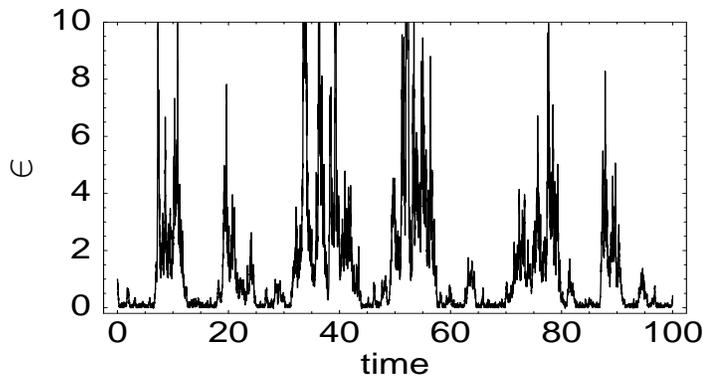}
   \caption{A sample of a time-series of $\varepsilon(t)$ when the pressure-gradient takes a value in the bi-stable regime. Bursts are observed intermittently. It means the transition between the thermal noise state and the turbulent state occurs stochastically.}
\end{figure}

When the value of the pressure-gradient is larger than that of the linear stability boundary (see Fig. (4)), bursts are always observed. It means that the system is always in the turbulent state.
\begin{figure}
   \includegraphics[width=10cm, height=5cm]{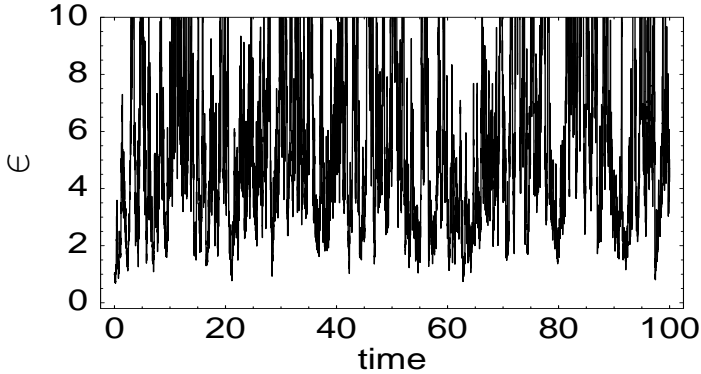}
   \caption{A sample of a time-series of $\varepsilon(t)$ when the value of the pressure-gradient is larger than that of the linear stability boundary. Bursts occurs simultaneously since the system in the turbulent state.}
\end{figure}

\section{The Transition Rates}
In order to formulate the above stochastic transition phenomena in the
bi-stable regime, we introduce the transition rates. There are
transitions in two opposite direction: the transition from the thermal
noise state to the turbulent noise state, which we call ``the forward
transition, and the transition in the opposite direction is called ``the
backward transition''. There are two transition rates. One is the forward transition rates $r_f$ which is the frequency of occurrence of the forward transition per unit time and the other is the backward transition rate $r_b$ defined similarly as the frequency of occurrence of the backward transition per unit time. 

It is important to note that these quantities are observable quantities. It is easily shown that the forward transition rate is equal to the average of inverse of the residence time at the thermal noise state and the backward transition rate is equal to the average of inverse of the residence time at the turbulent state. Therefore, these transition rates can be measured from the time serieses of fluctuation. 

We analyze in which region of the value of the pressure-gradient the
transition occurs frequently. The transition rates are calculated with
the formulae derived in \cite{Itoh00}. Two figures, Fig. (5) and Fig. (6), show the pressure-gradient dependence of the forward transition rate and the backward transition rate in the bi-stable regime. 
\begin{figure}[hptd]
   \includegraphics[width=12cm,height=8cm,keepaspectratio]{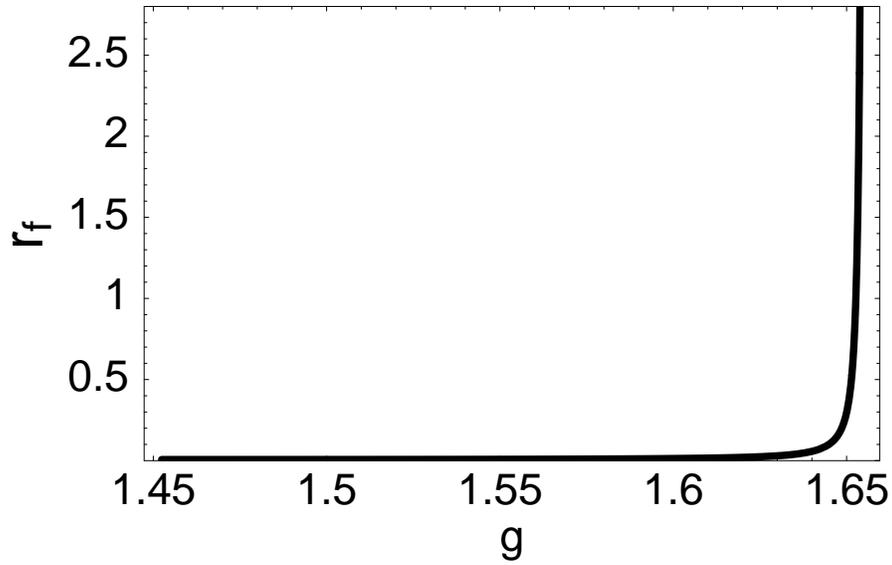}
   \caption{The pressure-gradient ($g$) dependence of the forward transition rate per unit time in the bi-stable regime. The left edge and the right edge of the horizontal axis corresponds to the subcritical bifurcation point and the linear stability boundary. It is seen that the forward transition occurs mainly in the vicinity of the linear stability boundary.}
\end{figure}
\begin{figure}[hptd]
   \includegraphics[width=12cm,height=8cm,keepaspectratio]{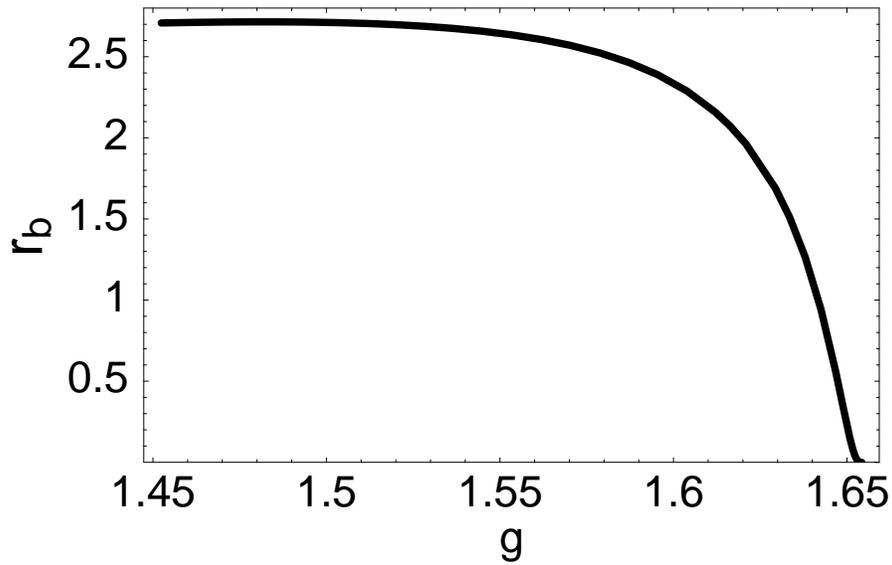}
   \caption{The pressure-gradient dependence of the backward transition rate per unit time in the bi-stable regime. It is seen that the backward transition occurs in the almost entire bi-stable regime.}
\end{figure}

The forward transition triggered by the thermal noise occurs mainly in
the vicinity of the linear stability boundary. In contrast, it
is clearly seen that the backward transition occurs in the almost entire
bi-stable regime. This behavior is due to strong turbulent
fluctuation. It is noted that the backward transition, i.e. the
transition in a turbulence, occurs in a ``region'' instead of a ``point''.

\section{Summary and Discussion}
Summarizing our work, we applied the statistical theory of plasma
turbulence to problems of the transition phenomena of submarginal
turbulence. By numerically solving the Langevin equation, the typical
time-development of fluctuation is obtained. It tells that the
transition for the model of inhomogeneous plasma occurs stochastically
and suggests how the transition phenomena due to subcritical bifurcation
may look in time-serieses obtained in real experiments. 

Furthermore, we
obtained the pressure-gradient dependence of the transition rates. It is
shown that the backward transition occur with almost equal frequency in
the entire bi-stable regime, so the transition occurs in a
``region''. 
The concept ``transition region'' is necessary in the analysis of data 
obtained by real experiments.


It is important to discuss whether the transition phenomena 
considered in this paper can be observed in real experiments. 
Since the characteristic time-scale of the 
transition is given by the inverse of the transition rate, 
observability depends on the interrelation between the time resolution of 
observation $\triangle t$ and the transition rate $r$. 
When $\triangle t$ is much smaller than $1/r$, the transition phenomena 
as shown in Fig. (3) are expected to be observed. On the other hand, 
when $\triangle t$ is of the same order of $1/r$ or larger than $1/r$,  
transition phenomena average out and 
only the average over $\triangle t$ is observed.
This discussion is generic regardless of the type of transition, e.g. 
the transition between the thermal noise state and the turbulent state, 
L/H transition etc.

\end{document}